\documentstyle[preprint,floats,aps,epsf,prb]{revtex}
\def\c60{$\rm C_{60}$}
\def\etal{{\it et al.}}
\begin{document}
\draft

\title{On The Vibrational Modes of $\rm C_{60}$}

\author{Dennis P. Clougherty\cite{byline} and John P. Gorman}

\address{Department of Physics, University of Vermont, Burlington, VT 
05405}

\date{January 16, 1995}

\maketitle

\begin{abstract}

The vibrational spectrum of \c60 is compared to the spectrum of 
 a classical isotropic elastic spherical shell. 
We show correlations between the low frequency
 modes of \c60 and those of the spherical shell.
We find the spherical model gives the approximate frequency
ordering for the low frequency modes. 
We estimate a Poisson ratio of $\sigma\approx 0.30$ and a transverse
speed of sound of $v_s\approx 1800$ m/s for the equivalent elastic shell.
We also find that 
$\omega({\rm M_1})/\omega({\rm M_0})=\sqrt{3\over 2}$
for the shell modes ${\rm M_0}$ and ${\rm M_1}$, independent of
elastic constants. We find that this ratio compares favorably with an
experimental value of $1.17$. 
\end{abstract} 


\section{Introduction}
\label{sec:intro}

The superconductivity of \c60-derived solids owes its
origin in part to features of 
the vibrational spectrum\ \onlinecite{johnson,varma,schluter} 
of \c60.  Consequently, a microscopic understanding of 
superconductivity requires an understanding of the 
vibrational modes. While there have previously been detailed computational 
studies\ \cite{wu,negri,stanton,weeks,adams,jishi} yielding vibrational
frequencies in excellent 
agreement with experiment, additional insight into the nature of modes
can be obtained by comparing the vibrational spectrum of
quasi-spherical \c60 with that of its continuum analog: a thin, elastic,
spherical shell. 

The electronic structure of \c60 has been previously described in a
similiar fashion, where \c60 is thought of roughly as
a spherical system with continuous rotational symmetry, invariant
under the group $\rm O(3)$. This leads to electronic multiplets,
labeled by $\ell$ with degeneracy $2\ell+1$. The true molecule has
only discrete icosahedral ($\rm I_h$) symmetry; and, as is commonly
done in crystal field theory, a correspondence can be made between the
multiplets under the approximate high symmetry group and their
descendants under the lower symmetry.

This correspondence between orbitals under $\rm O(3)$ and orbitals under
$\rm I_h$ was made by Haddon \etal\ \cite{haddon}
The $\rm O(3)$ multiplets order with energy increasing
with increasing $\ell$, implying that the energy scale associated with
changes in radial quantum number, $\epsilon_r$, is much larger than
the rotational energy scale, $\hbar^2\over 2mR^2$, where $R$ is the
length scale of the spherical potential and $m$ is the electron mass.
The most energetic of the 60 $\pi$-electrons therefore occupy the $\ell=5$
multiplet. 

Under the lower symmetry $\rm I_h$, multiplets of the spherical model
for $\ell > 2$ are split under $\rm I_h$, as icosahedral symmetry can
not support an orbital degeneracy larger than 5. For the occupied
orbitals, these gaps are smaller than the spacing between adjacent
$\ell$ multiplets; consequently, the correlation between $\ell$
multiplets and nearly degenerate manifolds in \c60 is clear.

We expect the comparison of vibrations of an elastic shell to that of
the \c60 molecule to give qualitative insight, but we certainly do not
expect close quantitative agreement. The continuum elastic
approximation is only justified for the lowest frequency modes such that
${\omega}<<\frac{v_s}{d}$, where $d$ is a typical nearest-neighbor
distance, and $v_s$ is the transverse speed of sound on the shell.

Furthermore, the linear elastic approximation assumes only local
interactions. It is known that for force constant models to get good
agreement with the experimentally measured vibrations of \c60, it is
necessary to include up to third-nearest neighbor interactions.\cite{jishi} 
Hence, a linear elastic model would not expect to yield good
quantitative agreement with experiment over the entire spectrum.
Nevertheless, to deepen our understanding of the character of the
complex modes of \c60, we find it useful to compare the molecular
modes with those of an elastic spherical shell, where the eigenmodes
and eigenfrequencies can be described by compact analytic expressions.

\section{Modes of the Thin Spherical Shell}
\label{sec:shell}

We summarize the results obtained by Lamb\ \cite{lamb} concerning the
vibrations of a thin spherical shell. The modes
can be divided into two classes. Modes of the first class have
displacements which are 
purely tangential to the surface of the shell. We denote the $\ell$th
of these shear modes by $\rm S_\ell$. The frequency spectrum of shear
modes is given by 

\begin{equation}
\Omega^{(S)}_\ell=\sqrt{(\ell-1)(\ell+2)}
\label{shearf}
\end{equation}
where $\Omega$ is the dimensionless frequency, 
\begin{equation}
\Omega={\omega R\over v_s},
\end{equation}
$R$ is the radius of the shell, and $v_s$ is the transverse speed of
sound on the shell. We note that $\Omega$ monotonically increases with
increasing $\ell$, and the lowest frequency shear mode has $\ell=2$.

The modes of the second class have displacements which have radial
character. Of these mixed modes, the $\ell=0$ mode is purely radial.
Its dimensionless frequency is given by
\begin{equation}
\Omega^{(M)}_0=2\sqrt{\gamma}
\label{m0}
\end{equation}
where $\gamma$ is related to the Poisson ratio of the shell $\sigma$,
\begin{equation}
\gamma={1+\sigma\over 1-\sigma}
\end{equation}
the Poisson ratio being the dimensionless 
ratio of transverse compression to longitudinal
extension.

The mixed modes with $\ell>0$ have displacements with both radial and
tangential character. The dimensionless frequency of the $\ell=1$ mode
is
\begin{equation}
\Omega^{(M)}_1=\sqrt{6\gamma}
\label{m1}
\end{equation}

\begin{figure}
\protect\centerline{\epsfxsize=6in \epsfbox{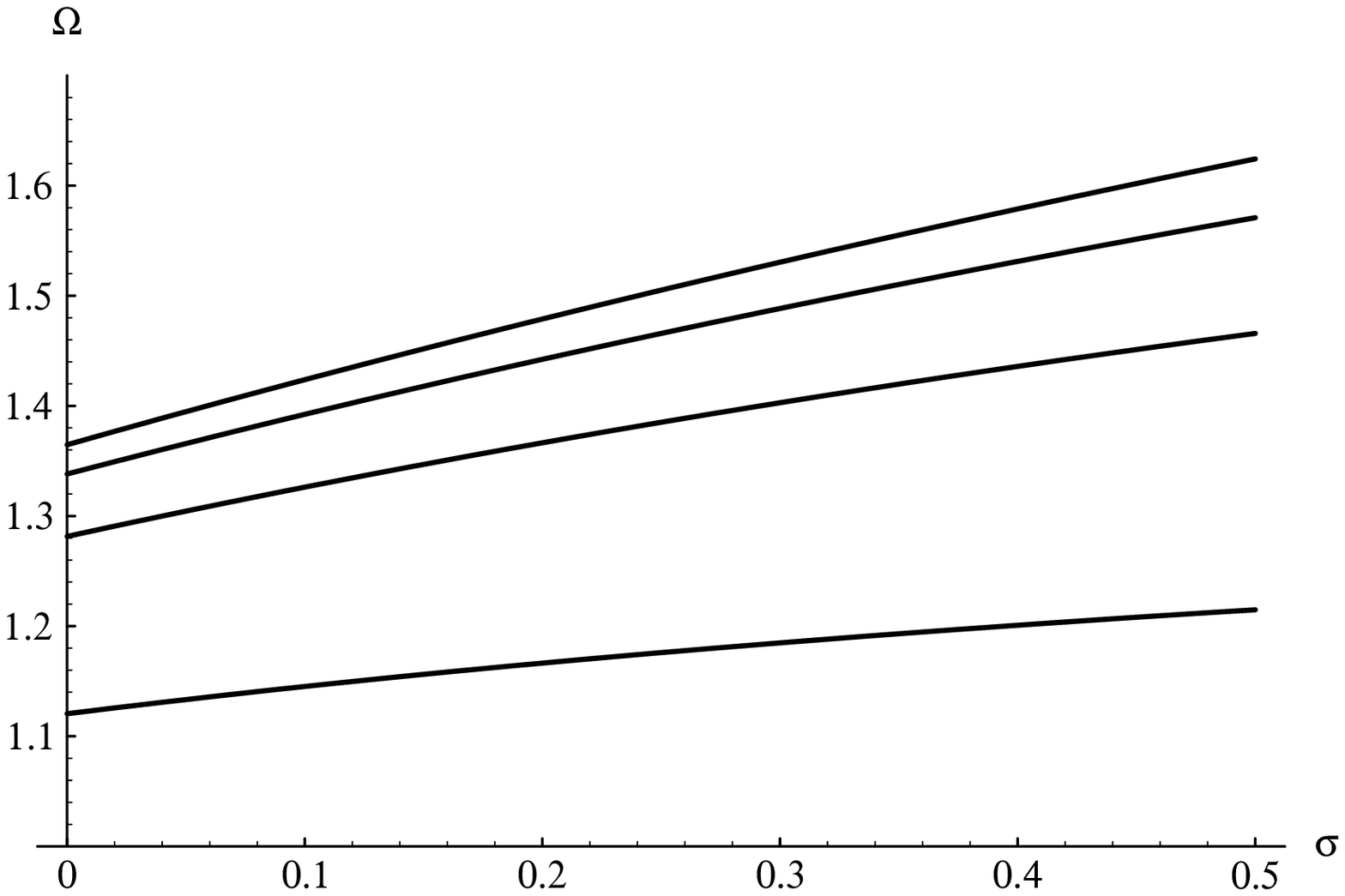}}
\protect\caption{Lowest frequency mixed modes $\rm M_{\ell-}$ ($\ell=$2, 3,
and 4) of a thin elastic shell vs Poisson Ratio $\sigma$. (Frequencies
are in dimensionless units of $v_s/R$.)
\protect\label{hy2}}
\end{figure}

\begin{figure}
\protect\centerline{\epsfxsize=6in \epsfbox{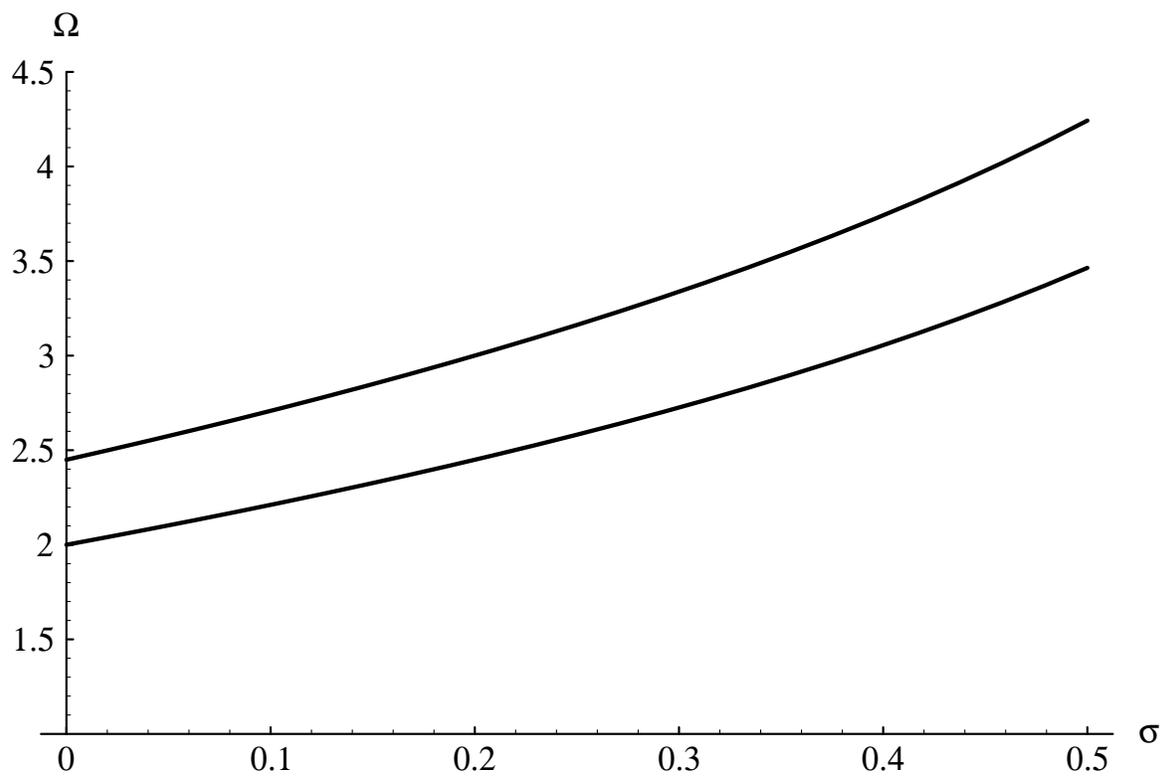}}
\protect\caption{Frequency of the $\rm M_{0}$ and $\rm M_{1}$ modes of a
thin elastic shell vs Poisson Ratio $\sigma$. (Frequencies are in
dimensionless units of $v_s/R$.)
\protect\label{hy01}}
\end{figure}

In the case of mixed modes with $\ell\ge 2$, each frequency has two
branches whose frequencies are 
\begin{equation}
\Omega^{(M)}_{\ell\pm}=\bigg[{\frac{\nu}{2}\pm\frac{1}{2}\big(\nu^2
-16(\ell^2+\ell-2)\gamma\big)^{1\over 2}}\bigg]^{1\over 2}
\label{m-}
\end{equation}
where $\nu={(\ell^2+\ell+4)\gamma+\ell^2+\ell-2}$.
We denote the mixed modes by $\rm M_{\ell(\pm)}$. We plot selected
shell frequencies as a function of the Poisson ratio in Figs.\
\ref{hy2} and \ref{hy01}.

\section{Results and Discussion}

We establish correlations between the modes of the shell and the modes
of \c60. The breathing mode of \c60 ($\rm A_g(1)$) can be
unambiguously matched with its counterpart mode of the shell, 
the mixed mode $M_0$. Both modes have exclusively radial character and
transform as the trivial representation of their respective symmetry groups.

\c60 has another mode preserving the icosahedral symmetry, but with
pure tangential character\ \cite{pickett}. We note that the decomposition of
$\ell=6$ representation under $\rm I_h$ contains $\rm A_g$\ \cite{fowler}.
We conclude that the $\rm A_g(2)$ mode is predominantly a descendant
of the $\rm S_6$ shell mode. 

The lowest frequency mode of the shell, $\rm M_{2-}$, has a degeneracy
of 5. Under $\rm I_h$, we expect this degeneracy to still be
supported. Experimentally, the lowest frequency mode of \c60, the
so-called squashing mode $\rm H_g(1)$, has mixed displacement
character\ \cite{pickett} and the same degeneracy as the
$\rm M_{2-}$ mode on the shell; thus, we infer that they are correlated.

The next lowest frequency mode of the shell, $\rm M_{3-}$, has a
degeneracy of 7, which decomposes under $\rm I_h$ into $\rm T_{2u}$
and $\rm G_u$ modes. We correlate $\rm M_{3-}$ with the nearly
degenerate multiplet $\rm T_{2u}(1)$ and $\rm G_u(1)$. Similarly, the
$\rm H_g(2)$ and $\rm G_g(1)$ levels must have evolved primarily from
the splitting of the $\rm M_{4-}$ mode. 

As the $\rm M_{1}$ shell mode must remain unsplit, we correlate it with a
$\rm T_{1u}$ type mode under $\rm I_h$. There are two $\rm T_{1u}$
modes which are close in frequency. From stereographic
projections\ \cite{weeks} of the modes, we conclude that 
$\rm T_{1u}(2)$ is primarily a descendant of $\rm M_{1}$, while $\rm
T_{1u}(1)$ results from the splitting of the $\rm M_{5-}$ mode.

\begin{figure}
\protect\centerline{\epsfxsize=6in \epsfbox{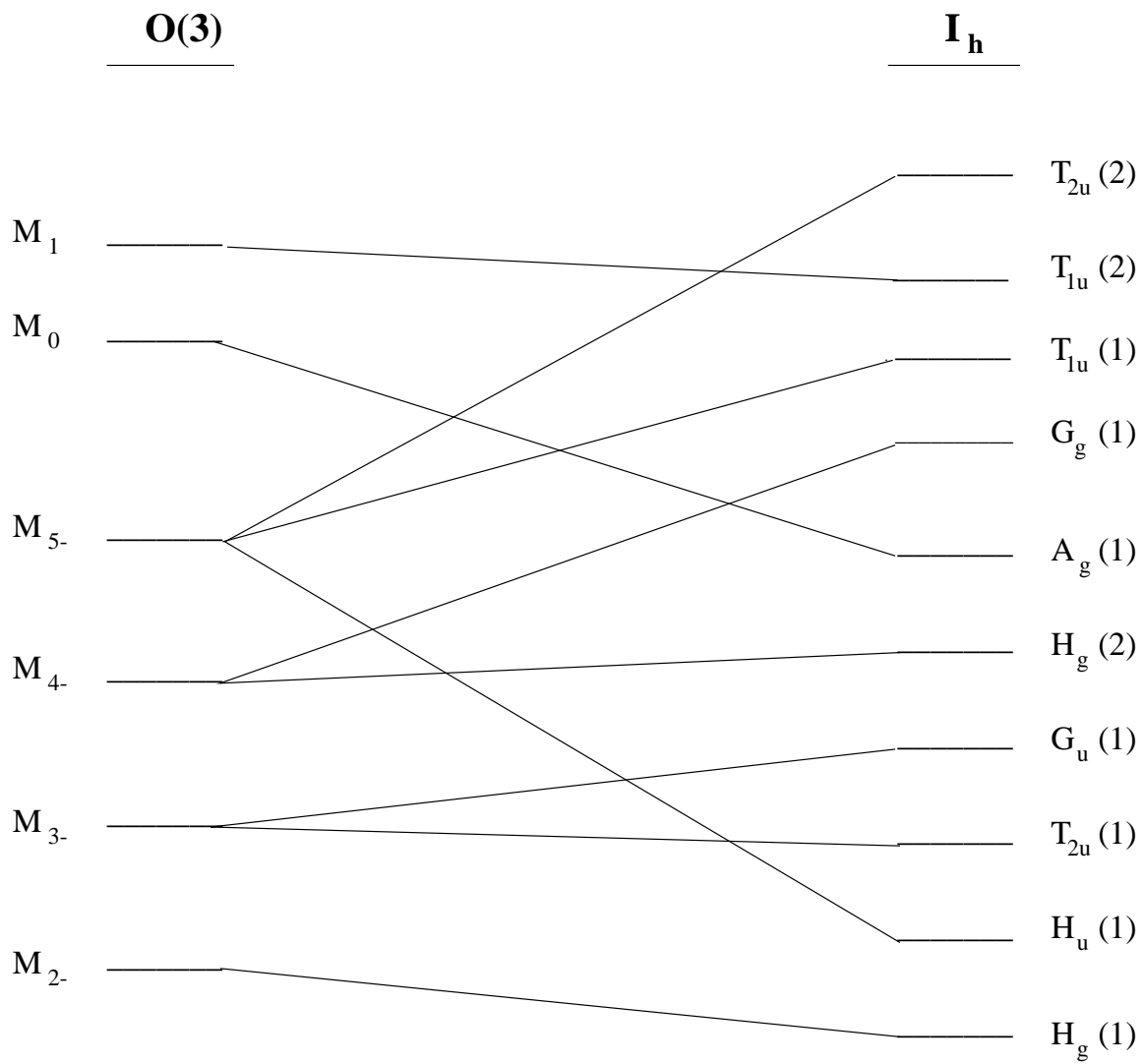}}
\protect\bigskip
\protect\caption{Correlation diagram of selected low frequency modes of a thin
elastic shell with vibrational modes of \c60.
\protect\label{corr}}
\end{figure}

For $\ell\ge 5$, mixed shell modes have
small spacings compared to the splittings under $\rm I_h$. There are
no real predominant ``progenitor'' levels, as mixing among many shell
modes must be involved in determining the final character of the
displacements.
We summarize our low frequency correlations in Fig.\ \ref{corr}. 

The spectrum of the shell results from one structural parameter 
and two elastic parameters. We adjust the elastic parameters to fit to
the low frequency portion of the spectrum. We use a value of 3.56\ \AA
\ for $R$, as obtained from experiment. 
We estimate the following parameter values from a fit to the
$\rm A_g(1)$ and $\rm H_u(1)$ modes using Eqs.\ \ref{m0} and \ref{shearf}:
$\sigma\approx0.30$ and 
$v_s\approx1800$ m/s. We choose to fit to these modes as their
correlation with shell modes is clear, and they stem from multiplets
whose degeneracy is preserved under $\rm I_h$. 

Combining Eqs.\ \ref{m0} and \ref{m1}, we observe that 
\begin{equation}
\frac{\omega({\rm M_1})}{\omega({\rm M_0})}=\sqrt{3\over 2}\approx 1.22
\end{equation}
{\em independent} of elastic parameters. Based on our correlation
diagram, we can compare this ratio to experiment. We use the 
value of 493 cm$^{-1}$ for ${\omega({\rm A_g(1)})}$\ \cite{zhou}, and 
a value of 577 cm$^{-1}$ for ${\omega({\rm T_{1u}(2)})}$\
\cite{huffman} to obtain a ratio of 
${\omega({\rm T_{1u}(2)})}/{\omega({\rm A_g(1)})}= 1.17$. 

The
reasonable agreement with ${\omega({\rm M_1})}/{\omega({\rm M_0})}$
gives support to the interpretation of these modes as progenitors of 
the ${\rm A_g(1)}$ and ${\rm T_{1u}(2)}$ modes. 
A comparison of selected spherical 
model frequencies to experimental frequencies is given in
Table I.
Perhaps the real utility of
the correlations made is that
the known analytic forms of the eigenmodes might be used to 
approximate displacement fields in calculations.

\acknowledgments

We thank F.~Anderson, W.~Leenstra, and J.~Wu for valuable
discussions. Acknowledgment is made to the Donors of The Petroleum
Research Fund, administered by the American Chemical Society, for
support of this research.

\vfil\eject

\begin{table}
\caption{Comparison of selected 
vibrational frequencies of sphere, $\omega$, with the experimental 
vibrational frequencies
\c60,
$\omega_{\rm exp}$.}
\begin{tabular}{lrl}
Mode&$\omega$&$\omega_{\rm exp}$$^{\rm a}$\\
\tableline
M$_0$&493$^*$&493\\
M$_1$&577&589\\
M$_{2-}$&214&269\\
M$_{3-}$&253&367, 385\\
S$_{2}$&361$^*$&361\\
S$_{3}$&571&498, 541\\
S$_{4}$&766&801, 929\\
\end{tabular}
\label{radii}
{$^{\rm a}$\ Refs.\ \cite{jishi,zhou}}

{$^{*}$ Fit to experiment}

\end{table}

\end{document}